\documentclass[a4paper,12pt]{article}

\usepackage{amsmath}
\usepackage[psamsfonts]{amssymb}
\usepackage{rsfs}
\usepackage{bm}

\usepackage{cite}
\usepackage[dvips]{graphicx}
\usepackage{color}

\begin{document} 

\begin{titlepage}

\baselineskip 10pt
\hrule 
\vskip 5pt
\leftline{}
\leftline{Chiba Univ. Preprint
          \hfill   \small \hbox{\bf CHIBA-EP-140}}
\leftline{\hfill   \small \hbox{hep-th/0306195}}
\leftline{\hfill   \small \hbox{June 2003}}
\vskip 5pt
\baselineskip 14pt
\hrule 
\vskip 1.0cm
\centerline{\Large\bf 
A physical meaning of 
} 
\vskip 0.3cm
\centerline{\Large\bf  
  mixed gluon--ghost condensate  of  
}
\vskip 0.3cm
\centerline{\Large\bf  
mass dimension two }
\vskip 0.3cm
\centerline{\large\bf  
}

\vskip 0.5cm

\centerline{{\bf 
Kei-Ichi Kondo$^{\dagger,{1}}$ 
}}  
\vskip 0.5cm
\centerline{\it
${}^{\dagger}$Department of Physics, Faculty of Science, 
Chiba University, Chiba 263-8522, Japan
}
\vskip 1cm

\begin{abstract}
We demonstrate that a clear physical content and relevance can be attributed to the  on-shell BRST-invariant mixed gluon--ghost condensate of mass dimension two which was recently proposed by the author.
We argue that a gauge invariant observable is associated with the mixed condensate.

\end{abstract}

Key words:  vacuum condensate, mass gap, confinement,  Yang--Mills theory, 

PACS: 12.38.Aw, 12.38.Lg 
\hrule  
\vskip 0.1cm
${}^1$ 
  E-mail:  {\tt kondok@faculty.chiba-u.jp}

\par 
\par\noindent


\vskip 0.5cm

\newpage

\vskip 0.5cm  



\end{titlepage}


\pagenumbering{arabic}

\baselineskip 14pt

Recently, a novel vacuum condensate has been proposed \cite{Kondo01} as the vacuum expectation value (VEV) of the composite operator $\mathcal{O}$ of mass dimension two where $\mathcal{O}$ is on-shell Becchi-Rouet-Stora-Tyutin (BRST) invariant operator:
\begin{align}
  \mathcal{O} := \Omega^{-1} \int d^4x \ {\rm tr}_{G/H} \left[ {1 \over 2} \mathscr{A}_\mu \mathscr{A}^\mu + \lambda i \bar{\mathscr{C}} \mathscr{C} \right] ,
\label{mixed}
\end{align}
where $\Omega$ is the volume of the spacetime and all the fields $\Phi:=\{ \mathscr{A}_\mu, \mathscr{C}, \bar{\mathscr{C}} \}$ take values in the Lie algebra  of the gauge group $G$, i.e, $\Phi=\Phi^A T^A$ with generators $T^A$ ($A=1, \cdots, \text{dim}G$).  Here the ${\rm tr}_{G/H}$ means the trace over the broken generators when the original gauge group $G$ is broken to the subgroup $H$ by the (partial) gauge fixing and $\lambda$ is the gauge fixing parameter.  
In the Lorentz gauge, $G$ is completely broken, i.e., $H=\{ 0 \}$. In the Maximal Abelian (MA) gauge, $H$ is the maximal torus subgroup of $G$, i.e., $H=U(1)^{N-1}$ for $G=SU(N)$.  
Here the {\it on-shell} BRST is used to mean that the BRST transformation is defined by eliminating the Nakanishi-Lautrup (NL) auxiliary field $\mathscr{B}$ and the BRST algebra is closed among the gauge field $\mathscr{A}_\mu$, ghost field $\mathscr{C}$ and antighost field $\bar{\mathscr{C}}$ without the NL field $\mathscr{B}$.   It is important to keep the fact in mind that the auxiliary field $\mathscr{B}$ does not have the kinetic term.  

In this paper we point out a transparent physical meaning of the (on-shell) BRST-invariant composite operator $\mathcal{O}$ of mass dimension two and its vacuum expectation value, $\langle \mathcal{O} \rangle$, i.e., the mixed vacuum condensate of mass dimension two.  
The on-shell is used to mean that the NL field $\mathscr{B}$ is eliminated by using the equation of motion (EOM) for $\mathscr{B}$ or it is integrated out by Gaussian integration in the functional integration. 
In the on-shell BRST transformation $\bm{\delta}_{\rm os}$, the EOMs for the fields $\mathscr{A}_\mu, \mathscr{C}, \bar{\mathscr{C}}$ are not assumed to hold, except for $\mathscr{B}$. 
It is easy to show that the action obtained after eliminating the NL field $\mathscr{B}$ is invariant under the on-shell BRST transformation. 
Therefore, the on-shell BRST transformation $\bm{\delta}_{\rm os}$ plays the same role in the $\mathscr{B}$-eliminated action as the off-shell BRST transformation $\bm{\delta}_{\rm B}$ in the $\mathscr{B}$-included action. 
An inconvenience of the on-shell BRST transformation lies in a fact  that the nilpotency is not fully maintained, e.g., the nilpotency of the on-shell BRST transformation for ghost (antighost) is satisfied only when the EOM for the antighost (ghost) is used.

An apparent advantage of $\mathcal{O}$ is to be able to incorporate simultaneously two novel vacuum condensates, each of which has  independently been proposed  recently; 
the gluon pair condensation 
$\langle \mathscr{A}_\mu^A  \mathscr{A}_\mu^A \rangle$ 
in the Lorentz covariant gauge \cite{Boucaudetal00,GSZ01} (see \cite{LS88} for the old works) and the ghost--antighost condensation 
$\langle \bar{C}^a C^a \rangle$ 
in the MA gauge \cite{Schaden99,KS00} where the index $a$ runs over the off-diagonal components only. 
In particular, 
$\langle \mathcal{O} \rangle$ seems to reduce to the gluon condensation 
$\langle \mathscr{A}_\mu^A \mathscr{A}_\mu^A \rangle$  
in the limit $\lambda \rightarrow 0$ of Landau gauge.

Quite recently, the physical meaning of $\mathcal{O}$ was seriously re-examined \cite{Gripaios03,EF03} based on the general theory of the BRST cohomology \cite{BRSTcohomology}. 
The purpose of this paper is to answer some of the criticisms raised there without using the abstract theory of BRST cohomology by pointing out a transparent physical meaning of $\mathcal{O}$. 

In this paper we argue 
\begin{enumerate}
\item[1.] 
The requirement of on-shell BRST invariance for $\mathcal{O}$, i.e,  $\bm{\delta}_{\rm os}\mathcal{O}=0$ is equivalent to the gauge invariance for $\mathcal{O}'$, i.e, $\bm{\delta}_{\rm \omega}\mathcal{O}'=0$. 

\item[2.] The VEV of $\mathcal{O}$ is equivalent to the VEV of the gauge invariant operator $\mathcal{O}'$ which has zero ghost number and is written in terms of the gauge field alone, 
$\langle \mathcal{O} \rangle = \langle \mathcal{O}' \rangle$. 
Therefore, 
$\langle \mathcal{O} \rangle \not= 0$
is a gauge invariant and gauge independent statement. 

\item[3.]
 The gauge invariant operator $\mathcal{O}'$ is nonlocal and can be non-linear in the non-Abelian gauge theory.

\end{enumerate}
By {\it local}, it is meant that the functional depends on the fields and a finite number of their derivatives all of which are evaluated at the same point in spacetime.  
In this sense, the transverse and longitudinal modes of the gauge field are nonlocal objects. 

Moreover, the operator $\mathcal{O}$ is invariant under the SL(2,R) transformation generated by the BRST charge $Q_B$, the anti-BRST charge $\bar{Q}_B$  and the ghost number charge $Q_c$ which constitute the extended Becchi-Rouet-Stora-Nakanishi-Ojima (BRSNO) algebra \cite{NO80} in the generalized Lorentz gauge \cite{CF76} and the modified MA gauge \cite{KondoII,Schaden99,KS00}.

In order to illustrate the validity of the above claims, we begin with the free Abelian gauge theory.  
In the Abelian gauge theory, the off-shell BRST and anti-BRST transformations are given by 
\begin{subequations}
\begin{align}
 \bm{\delta}_{\rm B} A_\mu(x) &=
 \partial_\mu C(x),  \\
 \bm{\delta}_{\rm B} C(x) &=0 ,
\\
 \bm{\delta}_{\rm B} \bar{C}(x) &=i B(x) ,
\\
 \bm{\delta}_{\rm B} B(x) &=0 ,
\label{BRSTA1}
\end{align}
\end{subequations}
and
\begin{subequations}
\begin{align}
 \bar{\bm{\delta}}_{\rm B} A_\mu(x) &=
\partial_\mu \bar{C}(x)  , \\
 \bar{\bm{\delta}}_{\rm B} \bar{C}(x) &=0 ,
 \\
 \bar{\bm{\delta}}_{\rm B} C(x) &=i \bar{B}(x) ,
\\
 \bar{\bm{\delta}}_{\rm B} \bar{B}(x) &=0 ,
\label{BRSTA2}
\end{align}
\end{subequations}
where $\bar{B}$ is defined by
\begin{equation}
  \bar{B}(x) =-B(x)  .
\end{equation}
Here 
$A_\mu, B, C$ and $\bar{C}$ are the Abelian gauge field, the NL auxiliary field, the FP ghost and antighost fields respectively.
The BRST and anti-BRST transformations are nilpotent, i.e., 
$\{ \bm{\delta}_{\rm B}, \bm{\delta}_{\rm B} \}= 0 = \{ \bar{\bm{\delta}}_{\rm B}, \bar{\bm{\delta}}_{\rm B}\}$ and they anti-commute, i.e., $\{ \bm{\delta}_{\rm B}, \bar{\bm{\delta}}_{\rm B}\}=0$. 

\par
We consider  the theory with the Lagrangian density
\begin{equation}
  \mathscr{L}^{\rm total} = -{1 \over 4}(\partial_\mu A_\nu - \partial_\nu A_\mu)^2 + \mathscr{L}_{\rm GF+FP} .
\end{equation}
 As pointed out in \cite{KondoIII}, the gauge-fixing (GF) and the Faddeev-Popov (FP) ghost term $\mathscr{L}_{\rm GF+FP}$ in the Lorentz covariant gauge is rewritten into the simultaneous BRST and anti-BRST exact form,
\begin{equation}
  \mathscr{L}_{\rm GF+FP} =i \bm{\delta}_{\rm B} \bar{\bm{\delta}}_{\rm B} \left( {1 \over 2} A_\mu A^\mu + {\lambda \over 2}i  \bar{C} C \right) .
\label{GFA1}
\end{equation}
In fact, this is cast into the form,
\begin{align}
  \mathscr{L}_{\rm GF+FP} =i \bm{\delta}_{\rm B}  \left(   (\bar{\bm{\delta}}_{\rm B} A^\mu) A_\mu  - {\lambda \over 2}i  \bar{C} \bar{\bm{\delta}}_{\rm B} C \right) 
= - i \bm{\delta}_{\rm B}  \left[  \bar{C}  \left( \partial^\mu A_\mu + {\lambda \over 2}   B \right) \right] ,
\end{align}
which agrees with the conventional form up to a total-derivative term,\begin{align}
  \mathscr{L}_{\rm GF+FP} = B \partial^\mu A_\mu + {\lambda \over 2}B^2 + i \bar{C} \partial^\mu \partial_\mu C .
\end{align}
  By virtue of nilpotency, it is trivial to see that $\bm{\delta}_{\rm B} \mathscr{L}_{\rm GF+FP}=0
= \bar{\bm{\delta}}_{\rm B}\mathscr{L}_{\rm GF+FP}$.

Defining $\partial^2:=\partial^\mu \partial_\mu$, we can write down the EOM's as 
\begin{align}
  \partial^2 A_\mu - \partial_\mu \partial^\nu A_\nu - \partial_\mu B = 0 ,
\label{A}
\\ 
\lambda B + \partial^\mu A_\mu = 0,
\label{B}
\\
  \partial^2 C = 0,
\\
  \partial^2 \bar{C} = 0 .
\end{align}
The nilpotency of the on-shell BRST transformation for the ghost (antighost) can be checked by making use of the EOM of antighost (ghost).  
Taking the derivative of (\ref{A}) yields 
\begin{equation}
  \partial^2 B = 0 .
\end{equation}
Substituting $B$ of (\ref{B}) into (\ref{A}) leads to 
\begin{equation}
 \partial^2 A_\mu -(1-\lambda) \partial_\mu B = 0 .
\end{equation}

\par
If the NL field $B$ is eliminated by performing the functional integration or by making use of the EOM, 
$\lambda B = - \partial^\mu A_\mu$, 
then we obtain
\begin{equation}
  \mathscr{L}_{\rm GF+FP}' =- {1 \over 2\lambda} (\partial^\mu  A_\mu)^2  + i \bar{C} \partial^\mu \partial_\mu C .
\label{GFA2}
\end{equation}
In the Abelian gauge theory, the ghost and antighost decouple from the theory.  But they are necessary for our purpose.  
The on-shell BRST transformation is given by
\begin{subequations}
\begin{align}
 \bm{\delta}_{\rm os} A_\mu(x) &=
 \partial_\mu C(x),  \\
 \bm{\delta}_{\rm os} C(x) &=0 ,
\\
 \bm{\delta}_{\rm os} \bar{C}(x) &=-{i \over \lambda} \partial^\mu  A_\mu (x) ,
\label{BRSTA3}
\end{align}
\end{subequations}
while the on-shell anti-BRST transformation is 
\begin{subequations}
\begin{align}
 \bar{\bm{\delta}}_{\rm os} A_\mu(x) &=
\partial_\mu \bar{C}(x)  , \\
 \bar{\bm{\delta}}_{\rm os} \bar{C}(x) &=0 ,
 \\
 \bar{\bm{\delta}}_{\rm os} C(x) &=+ {i \over \lambda} \partial^\mu  A_\mu (x) .
\label{BRSTA4}
\end{align}
\end{subequations}
It is easy to check that 
$\bm{\delta}_{\rm os} \mathscr{L}_{\rm GF+FP}'=0
= \bar{\bm{\delta}}_{\rm os}\mathscr{L}_{\rm GF+FP}'$.

When $\lambda=1$, one can write the usual three-dimensional momentum representation for the gauge field,
\begin{align}
  A_\mu(x) &=   \sum_{\sigma=\pm,L,S}  A_\mu^{(\sigma)}(x)
\nonumber\\&
= \int {d^3 \bm{k} \over \sqrt{(2\pi)^3 2|\bm{k}|}} 
\sum_{\sigma=\pm,L,S} \left[ a(\bm{k},\sigma) \epsilon_\mu^{(\sigma)}(\bm{k}) e^{-ikx} + a^\dagger(\bm{k},\sigma) \epsilon_\mu^{(\sigma)}{}^*(\bm{k}) e^{+ikx} \right] \Big|_{k^0=|\bm{k}|} ,
\label{Ad}
\end{align}
 since it obeys the d'Alembert equation $\partial^2 A_\mu(x)=0$. 
Here $\epsilon_\mu^{(\sigma)}(\bm{k}) $ is the polarization vector which obeys the orthogonality relation
$\epsilon_\mu^{(\sigma)}{}^*(\bm{k})\epsilon^{(\tau)}{}^{\mu}(\bm{k})=-\eta^{\sigma\tau}$
and the completeness relation
$
 \sum_{\sigma,\tau} \epsilon_\mu^{(\sigma)}(\bm{k}) \eta^{\sigma\tau} \epsilon_\nu^{(\tau)}{}^*(\bm{k})=-g_{\mu\nu}
$
with the metric matrix:
\begin{align}
  (\eta)^{\sigma\tau} := \begin{pmatrix}
  1 & 0 & 0 & 0 \\
  0 & 1 & 0 & 0 \\
  0 & 0 & 0 & 1 \\
  0 & 0 & 1 & 0 
  \end{pmatrix} . 
\end{align}
The indices $\sigma=\pm, L,S$ denote the two transverse, longitudinal and scalar modes respectively. 
Their explicit forms are given as
$
 \epsilon^{(L)}{}^\mu(\bm{k}) = -ik^\mu 
$, 
and
$
 \epsilon^{(S)}{}^\mu(\bm{k}) = i(|\bm{k}|,\bm{k})/(2|\bm{k}|^2) 
$.

When $\lambda\not=1$, $A_\mu$ becomes the dipole field, i.e., 
$(\partial^2)^2 A_\mu(x) = 0$.  This fact prevents one from writing the usual three-dimensional momentum representation. But the four-dimensional momentum representation is possible, see \cite{Nakanishi75,NO90}, and the following argument can be extended to the $\lambda\not=1$ case. 

In what follows, therefore, we restrict our argument to the Feynman gauge $\lambda=1$ for simplicity. 
The ghost and antighost fields have the representation:
\begin{align}
  \binom{\bar{c}(x)}{c(x)} =   
 \int {d^3 \bm{k} \over \sqrt{(2\pi)^3 2|\bm{k}|}} 
 \left[ \binom{\bar{c}(\bm{k})}{c(\bm{k})} e^{-ikx} + \binom{\bar{c}^\dagger(\bm{k})}{c^\dagger(\bm{k})}  e^{+ikx} \right] \Big|_{k^0=|\bm{k}|} .
\label{Cd}
\end{align}

We show that the on-shell BRST closed operator $\mathcal{O}$ ($\bm{\delta}_{\rm os}\mathcal{O}=0$) is written as a sum of the gauge-invariant (but nonlocal) operator $\mathcal{O}'$ ($\bm{\delta}_{\omega}\mathcal{O}'=0$) and the on-shell BRST exact part:
\begin{align}
  \mathcal{O} = \mathcal{O}' + \bm{\delta}_{\rm os} \mathcal{O}'' ,
\label{decomp}
\end{align}
where
\begin{align}
  \mathcal{O}' = \Omega^{-1} 
 \int {d^3 \bm{k} \over (2\pi)^3 2|\bm{k}|} \sum_{\sigma=\pm}  a(\bm{k},\sigma) a^\dagger(\bm{k},\sigma)  ,
\end{align}
and
\begin{align}
  \mathcal{O}'' =  \Omega^{-1} 
 \int {d^3 \bm{k} \over (2\pi)^3 2|\bm{k}|} \lambda [i \bar{c}(\bm{k}) a^\dagger(\bm{k},L)+ \text{h.c.}] .
\end{align}
The on-shell BRST closedness 
$\bm{\delta}_{\rm os}\mathcal{O}=0$
follows from the nilpotency of on-shell BRST transformation,  
$\bm{\delta}_{\rm os}\bm{\delta}_{\rm os}=0$, since
$\bm{\delta}_{\omega}\mathcal{O}'=0$   implies 
$\bm{\delta}_{\rm os}\mathcal{O}'=0$. 
It is easy to see that eq.~(\ref{decomp}) is a decomposition of $\mathcal{O}$ into the physical and unphysical parts, since   
 the second term is written as
\begin{align}
   \bm{\delta}_{\rm os} \mathcal{O}'' 
= \Omega^{-1} 
 \int {d^3 \bm{k} \over (2\pi)^3 2|\bm{k}|} [a(\bm{k},S) a^\dagger(\bm{k},L)+ \lambda i \bar{c}(\bm{k}) c^\dagger(\bm{k}) + \text{h.c.}] ,
\end{align}
where we have used the orthogonality relation of the polarization vector and  the BRST transformation of the creation and annihilation operators:
\begin{subequations}
\begin{align}
 \bm{\delta}_{\rm os} a(\bm{k},T) &=  0 ,  \\
 \bm{\delta}_{\rm os} a(\bm{k},S) &=  0 ,  \\
 \bm{\delta}_{\rm os} a(\bm{k},L) &=   c(\bm{k}) ,  \\
 \bm{\delta}_{\rm os} c(\bm{k}) &=0 ,
\\
 \bm{\delta}_{\rm os} \bar{c}(\bm{k}) &=i B(\bm{k})= - i \lambda^{-1}  a(\bm{k},S) .
\label{BRSTA5}
\end{align}
\end{subequations}
and $(-ik^\mu) \epsilon_\mu^{(\sigma)}(\bm{k})=\delta^{\sigma S}$ for deriving the last equation. 
In the above equations, $\lambda$ should be understood to be equal to one.

The main result (\ref{decomp}) follows from a straightforward calculation by substituting the mode decomposition (\ref{Ad}), (\ref{Cd}) into (\ref{mixed}) and making use of the commutation and anticommutation relations for the creation and annihilation operators:
\begin{subequations}
\begin{align}
  [a(\bm{k},\sigma), a^\dagger(\bm{k}',\tau) ] =& \eta^{\sigma\tau} \delta^3(\bm{k}-\bm{k}') ,
\\
 \{ c(\bm{k}), \bar{c}^\dagger(\bm{k}') \} =& - \{ \bar{c}(\bm{k}), c^\dagger(\bm{k}') \} =  i \delta^3(\bm{k}-\bm{k}') ,
\end{align}
\end{subequations}
where all the other commutators and anticommutators are vanishing. 

If we take the VEV of $\mathcal{O}$, it reduces to the VEV 
$\langle \mathcal{O}' \rangle$
of the gauge-invariant quantity written in terms of the transverse gauge boson, 
\begin{equation}
  \langle \mathcal{O} \rangle = \langle \mathcal{O}' \rangle ,
\end{equation}
and
\begin{equation}
  \mathcal{O}' = \Omega^{-1} \int d^4x \sum_{\sigma=\pm} {1 \over 2}[A_\mu^{(\sigma)}(x)]^2 ,
\end{equation}
since the BRST charge annihilates the vacuum, 
$Q_B |0\rangle =0$.
Hence the VEV $\langle \mathcal{O}' \rangle$ must be gauge independent. 
The gauge field can be decomposed into the transverse and the longitudinal parts, 
$\bm{A}(x)=\bm{A}^T(x)+\bm{A}^L(x)$
where 
$\nabla \cdot \bm{A}^T(x)=0$
and 
$\nabla \times \bm{A}^L(x)=0$. 
The transverse and the longitudinal parts have the nonlocal expressions in terms of the original gauge field,
\begin{subequations}
\begin{align}
  \bm{A}^T(x) = \left( 1 - \nabla {1 \over \Delta} \nabla \cdot \right) \bm{A}(x) 
= \bm{A}(x^0,\bm{x}) + \partial_i^x \int d^3 \bm{y} {1 \over 4\pi|\bm{x}-\bm{y}|} \partial_j^y A^j(x^0,\bm{y})  ,
\label{AT}
\end{align}
and
\begin{align}
  \bm{A}^L(x) = \nabla {1 \over \Delta} \nabla \cdot  \bm{A}(x) 
=  - \partial_i^x \int d^3 \bm{y} {1 \over 4\pi|\bm{x}-\bm{y}|} \partial_j^y A^j(x^0,\bm{y})  .
\label{AL}
\end{align}
\end{subequations}
Therefore, $[\bm{A}^T(x)]^2$ in the  integrand of $\mathcal{O}'$ is no longer the local quantity and can not be the mass term to be added to the Yang-Mills Lagrangian. 

In the pure Abelian gauge theory, the pair condensation does not occur, since there is no interaction between gauge bosons. 
In quantum electrodynamics (QED), such interaction is provided by the fermion loop. In quantum chromodynamics (QCD), the gluon self-interaction can do the same job. 
Suppose that the interaction can be incorporated by perturbation theory.  
In the asymptotic region, the above argument given for the Abelian gauge theory can be extended to the non-Abelian gauge theory simply by replacing the fields $\Phi(x) = \{ A_\mu(x), C(x), \bar{C}(x), B(x) \}$ in the Abelian gauge theory above with the asymptotic fields $\Phi_{as}^A(x)=\{ \mathscr{A}_\mu^A{}_{as}(x), \mathscr{C}^A_{as}(x), \bar{\mathscr{C}}^A_{as}(x), \mathscr{B}^A_{as}(x) \}$ with an adjoint index $A$ in the non-Abelian gauge theory, see \cite{KO79,Kugo89,NO90}. 
This framework is sufficient to discuss the scattering process in which the initial two transverse gauge bosons are scattered into the final two transverse gauge bosons. 
In fact, $a^A(\bm{k},L), a^A(\bm{k},S), c^A(\bm{k}), \bar{c}^A(\bm{k})$ are members of the quartet for each adjoint index $A$. 
Any quartet members can not be detected in the physical subspace $\mathcal{V}_{\rm phys}$ of positive semi-definite, because they can appear there only in zero-norm combination (quartet mechanism). 

\par
In the non-Abelian case, however, the explicit construction of the transverse mode is rather difficult, although it was attempted in the framework of perturbation theory up to the order $g^2$ \cite{LM94}:
\begin{align}
  \mathscr{A}_{\text{phys}}^i(x) = \left( \delta^{ij} - \partial^i {1 \over \Delta} \partial^j  \right) \Phi^j(x) , \quad 
\Phi^j = \Phi^j_{(0)} + g \Phi^j_{(1)} + g^2 \Phi^j_{(2)} + \cdots ,
\end{align}
which reduces to (\ref{AT}) in the Abelian limit. 
Here $\mathscr{A}_{\text{phys}}^i$ is transverse 
$\partial_i \mathscr{A}_{\text{phys}}^i(x)=0$ and 
 $\Phi$ can be obtained in such a way that $\mathscr{A}_{\text{phys}}^i$ is both 
BRST invariant, 
$\bm{\delta}_{\rm B} \mathscr{A}_{\text{phys}}^i(x)=0$, and gauge invariant, 
$\bm{\delta}_{\rm \omega} \mathscr{A}_{\text{phys}}^i(x)=0$, order by order of the coupling constant $g$.  The first two terms are  
$
 \Phi^j_{(0)}=\mathscr{A}^j  
$
and 
$
 \Phi^j_{(1)} = [v_{(1)}, \mathscr{A}^j] + {1 \over 2}[\partial^j v_{(1)}, v_{(1)}]  
$
where 
$
 v_{(1)} := {1 \over \Delta} \partial_j \mathscr{A}^j 
$,
see section 5 and Appendix A of \cite{LM94} for more details. 
Therefore, we can not prove the statement (\ref{decomp}) in the non-Abelian case, except for the asymptotic fields. 
Rather, it will be rather difficult to perform this program beyond the perturbation theory because of  the existence of the Gribov horizon, see \cite{BGI98,SBZ02} and references therein for  related works. 

According to the standard argument of BRST cohomology, there exists a one-to-one correspondence between the BRST cohomology at zero ghost number and the set of classical physical observables, i.e., gauge invariant functional of the gauge fields. 
We have shown that the classical physical observable to which the mixed vacuum condensate corresponds is the gauge-invariant dimension 2 operator, i.e., the transverse gauge boson pair, which is however nonlocal.
Therefore, an advantage of the mixed condensate proposed in \cite{Kondo01} is clear; The local and on-shell BRST invariant operator  avoids the nonlocality and plays in the gauge-fixed formulation  the same role as the gauge-invariant operator for any covariant gauge. 
The corresponding classical observable is invariant not only for the residual set of gauge transformation preserving the Lorentz condition, but also for the full gauge group $G$, contrary to the claim \cite{Gripaios03}.

Thus the physical meaning of the VEV 
$\langle \mathcal{O} \rangle$
is now clear;  
The VEV measures the transverse gluon pair condensation 
$\langle \mathcal{O}' \rangle$
due to attractive force generated by gluon self-interactions. 
Indeed, the ghost-antighost condensation is not BRST invariant, but ghost and antighost are  necessary to cancel the unphysical degrees of freedom corresponding to the longitudinal (spacelike) and scalar (timelike) modes of the gauge boson.
In this sense, ghost condensation is indispensable to convert the non-local operator $\mathcal{O}'$ to the local one $\mathcal{O}$ by adding unphysical degrees of freedom, which is easier to deal with. 
Incidentally, the operator $\mathcal{O}'$ is off-shell BRST closed. But the operator $\mathcal{O}$ can not be extendable to the off-shell BRST closed alternative, in agreement with \cite{EF03}. 

According to the theory of observables in Yang-Mills theory \cite{KO79,NO90,Kugo89}, $\mathscr{A}_{\text{phys}}^i$ 
is the observable in the strongest sense, i.e., 
$[ Q_B, \mathscr{A}_{\text{phys}}^i(x)]=0$.
This implies $Q_B (\mathscr{A}_{\text{phys}}^i(x) |0 \rangle) = 0$, i.e., $\mathscr{A}_{\text{phys}}^i(x) |0 \rangle \in \mathcal{V}_{\rm phys}$ for nonlocal operator $\mathscr{A}_{\text{phys}}^i(x)$. 
Therefore, the norm must be non-negative, i.e., 
$\langle 0| \mathscr{A}_{\text{phys}}^i(x)   \mathscr{A}_{\text{phys}}^i(x) |0 \rangle \ge 0$, which will lead to 
$\langle  \mathcal{O} \rangle \ge 0$.

Recently, the gluon pair condensation was measured on a lattice and it was found that the condensation was saturated by the contribution from the instanton configuration \cite{Boucaudetal02}.  
The instanton is the solution of the self-dual equation and  minimizes the Euclidean action.  It is a semi-classical configuration of the gauge field specified by the collective coordinates.  The solution obtained under the Ansatz 
\begin{align}
 \mathscr{A}_\mu^A(x) = \bar{\eta}^A_{\mu\nu} \partial_\nu f(x) 
\end{align}
using the (anti) self-dual 't Hooft tensor $\bar{\eta}^A_{\mu\nu}:=\epsilon_{A\mu\nu}-\delta_{A\mu}\delta_{\nu4}+\delta_{A\nu}\delta_{\mu4}$ is automatically transverse 
$\partial^\mu \mathscr{A}_\mu^A(x)=0$.  Therefore, the numerical result \cite{Boucaudetal02} is consistent with the claim in this paper. 
Within the effective potential for the composite operator to three loop order, on the other hand, the mixed condensate in the Landau gauge  $\lambda=0$ was calculated in the perturbation theory \cite{VKAV01,DVBG03} and it was attempted to extend to  arbitrary $\lambda$ \cite{DVLSS03}, provided that the vacuum satisfies the translation invariance.

Finally, it is straightforward to extend the above operator to the MA gauge \cite{Kondo01} and the general Lorentz non-covariant gauges \cite{BZ99} that include the Coulomb gauge as a limiting cases \cite{Zwanziger03}.

\section*{Acknowledgments}
The author would like to thank Atsushi Nakamura (Chiba Univ.) for valuable discussions on the BRST cohomology. 
This work is supported by Sumitomo Foundations, 
Grant-in-Aid for Scientific Research (C)14540243 from Japan Society for the Promotion of Science (JSPS), 
and in part by Grant-in-Aid for Scientific Research on Priority Areas (B)13135203 from
The Ministry of Education, Culture, Sports, Science and Technology (MEXT).


\end{document}